\documentstyle[bo99,epsfig]{article}

\title{The Energy Output of the Universe}

\author{A.C.~Fabian}                                                       
\affil{Institute of Astronomy, Madingley Road, Cambridge CB3 0HA, U.K. }      
\begin{document}

\maketitle

\begin{abstract}
The total energy emitted by the growth of massive black holes is large
and can be 10-50 per cent of that emitted by stars in the universe. I
show how the X-ray Background provides a good measure of this energy
and also why most accretion power is absorbed and re-emitted in the
far infrared band. A model for the obscured growth of massive black
holes is presented which accounts for both the high absorption and the
observed black hole to galaxy spheroid mass correlation. Future
missions should detect the obscured X-ray sources associated with the
growth of massive black holes.

\end{abstract}

\section{Introduction}

The X-ray Background (XRB) is the integrated emission from all X-ray
sources. Its hard spectrum has proved difficult to explain since, in
the 2--10~keV band, it is flat with a power-law of energy index 0.4.
This is flatter than the spectrum of any known common population of
objects. For the last decade the most popular explanation has been
that the XRB intensity is dominated by many absorbed sources (Setti \&
Woltjer 1989), with ranges of absorbing column density and redshift
causing the observed spectrum to be a power-law. The absorption model
has been extensively studied by Madau et al (1994), Matt \& Fabian
(1994), Comastri et al (1995), Celotti et al (1995), and Wilman \&
Fabian (1999). The most complete studies include Compton
down-scattering in the estimation of the observed spectrum of the
Compton-thick sources.

The absorption model is adopted here and is used in a simple way to
show that black holes grow by radiatively efficient accretion and to
determine a) the local mean density of black holes, b) the fraction of
accretion power which has been absorbed, and c) contraints on the
fraction of power in the Universe due to accretion (see also Fabian \&
Iwasawa 1999).  After some discussion of how so much obscuring
material can surround most sources, and how the nuclei might be
fuelled, I then outline a model of obscuration in a forming,
isothermal galaxy spheroid (Fabian 1999). The XRB is shown to be a key
diagnostic of the accretion power of the Universe.

\section{Accretion and the XRB}

I assume that the underlying active galactic nuclei (AGN) which power
the XRB have a quasar-like spectrum with an energy photon index of
one. The spectrum is then constant in a $\nu F_{\nu}$ sense (Fig.~1).
The action of photoelectric absorption by increasing amounts of
material, characterised by a column density $N_H$, is (Fig.~2) to cut
out the lower energy emission from the observed spectrum up to an
(approximate) energy $E\sim 10 N_H^{8/3}$~keV, where $N_H$ is in units
of $10^{24}$~cm$^{-2}$. As the column density exceeds about $1.5\times
10^{24}$~cm$^{-2}$ so the absorber becomes Compton thick and Compton
(electron) scattering causes the residual spectrum above this cutoff
to decrease in intensity. This means that the intensity observed above
about 30~keV is close to the intrinsic unattenuated intensity from
Compton-thin sources, and is a lower limit for Compton-thick ones.
Therefore the intensity of the XRB at 30~keV equals the normalization
of the XRB after correction for absorption by Compton-thin sources.
This normalization can be increased by a factor of $f^{-1}$ if only a
fraction $f$ of all the power emerges from sources which are Compton
thin.  $f$ is at most 3/4 (Maiolino et al 1998) and could be less than
one half.

The absorption-corrected XRB spectrum can then be extended into the
ultraviolet band assuming the mean quasar spectral energy distribution
of Elvis et al (1994). This shows that about 3 per cent of the power
from a typical quasar is emitted in the 2--10~keV band. The total
absorption-corrected AGN background can now be converted into an
energy density $\epsilon_{AGN}$ and thence, through the use of
$E=Mc^2$ or rather $\epsilon(1+\bar z)=\eta \rho c^2$ with an
accretion efficiency factor $\eta$ and a mean redshift $\bar z$ (since
photons lose energy in the expansion of the Universe but mass does
not), we have the mean density in black holes $\rho_{bh}$.

\begin{figure}
\centerline{\psfig{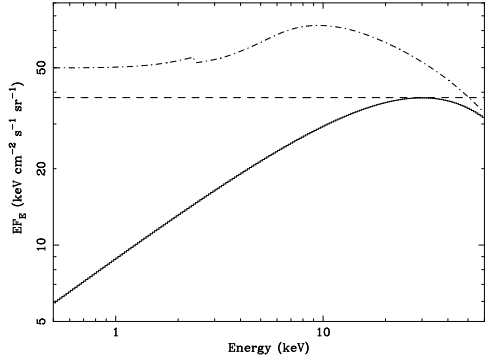}}
\caption{XRB spectrum (solid line) with the assumed unabsorbed
spectrum of photon index 2 (dotted line). A typical AGN spectrum with
reflection, in which the direct emission is a power law of photon
index 2 with an exponential cutoff of 300 keV is shown by the dot-dash
line, matching around the XRB peak. If unabsorbed quasars contribute
50 per cent of the XRB at 1 keV, then their contribution lies along
the bottom of the figure.}
\end{figure}

The resulting value of $\rho_{bh}= 6\times
10^5$~M$_{\odot}$~Mpc$^{-3}$ is about half the value found by
Magorrian et al (1998) from a study of ground-based optical data of
the cores of nearby galaxies, and in rough agreement with an HST
photometric study made by van der Marel (1999). Similar agreement has
been obtained by Salucci et al (1999) from a detailed considerations
of source counts etc. This close agreement emphasises that most of the
mass in black holes has been accreted by a radiatively efficient (but
obscured) process, and not by some inefficient process such as an
advective flow.  The correction required for absorption is extensive
and requires that most, about 85\%, of the accretion power has been
absorbed.

\section{AGN, the FIR Background and the energy from stars}

The absorbed power is assumed to be emitted in the Far Infrared (FIR)
bands, and when redshifted it should contribute to the sub-mm
background. The total predicted is about 3~nW~m$^{-2}$~sr$^{-1}$ which
is several tens percent of total the sub-mm background (Fixsen et al
1997; see also Almaini et al 1999 for estimates of the AGN
contribution to the sub-mm background). This suggest that to within a
factor of two the total integrated power (ie the total energy
released) from accretion onto black holes is about one quarter of that
from stars (mostly starlight but including supernovae),
i.e. $$E_{AGN}/E_{\star}\sim 0.25.$$ The details of any comparison
depend upon the history of the starlight and of the accretion. No
estimate of the contribution to the NIR and optical backgrounds, which
could lower the above value, has been made here.

A simple check on this is obtained from an argument due to G. Hasinger
(see Fabian \& Iwasawa 1999). Magorrian et al (1998) find the
following relation between the black hole mass $M_{bh}$ and spheroid
mass $M_{sph}$ of a galaxy: $$M_{bh}\approx 0.005 M_{sph},$$ so if the
total energy radiated $$E_{AGN}\approx 0.1 M_{bh}c^2$$ then
$$E_{AGN}\approx 0.1\times 0.005 M_{sph} c^2.$$ But the total energy
radiated by stars $$E_{\star}\approx 0.1\times 0.005 a^{-1} M_{sph}
c^2,$$ where the first term is the fraction of a star which undergoes
nuclear fusion and the second is the efficiency (in a $E=mc^2$ sense)
of that fusion. $a$ is the ratio of the present mass of the spheroid
to its original mass (many of the stars have evolved) and for a
Salpeter mass function is about 20 per cent. Therefore
$$E_{AGN}/E_{\star}\approx a \approx 0.2.$$

\section{Uncertainties}

The above estimate reduces to 0.1 if the scaling relation of van der
Marel (1999), which agrees better with the XRB intensity, is used, but
can increase towards unity if stellar mass loss is recycled into new
stars, so that $a\sim 1$. A mass-to-energy efficiency of $0.1$ has
been used but it can be $0.06$ if the black hole is not spinning. or
$0.42$ if it becomes a maximally spinning, Kerr, black hole. 

An even more extreme possibility which defines an upper limit on the
efficiency relative to the final (dead) black hole mass is to assume
that the black hole was maximally spinning during the accretion phase
and then spun down by, say, the Blandford-Znajek (1977) mechanism. The
total energy released relative to the final black hole mass allows for
an order of magnitude uncertainty in $\eta$ and thus $E_{AGN}$. Of
course a high value here, which maximises $E_{AGN}/E_{\star}$,
overpredicts the XRB intensity unless most of the growing phase of
black holes is Compton thick.  It is also possible that a significant
fraction of the power from an AGN is in the form of a wind and not
directly in radiation. As discussed later, growing black holes may be
both Compton thick and powering winds. If this is correct, then
$E_{AGN}/E_{\star}$ may be significantly higher than the estimate in
the last section.

\section{Obscuration, metallicity  and fuelling}

As outlined above, at least 85 per cent of accretion power is
absorbed. Since about ten per cent is in quasars which show very
little absorption, this means that most lines of sight out of the
remaining objects are highly absorbed. This is difficult for the
standard obscuring torus model, which could absorb perhaps one half to
two thirds of all sight lines. Even then it is unclear what inflates
the torus, which is supposed to be cold and molecular. Dissipation in
in a system of orbiting clouds should cause it to flatten into a disc,
with lowcovering factor.

Energy must be continuously injected into any cold absorbing cloud
system to keep it inflated and so sky covering. One plausible solution
is that a gas-rich star cluster surrounds the black hole and it is the
massive stars (winds and supernovae) which supply the energy (Fabian
et al 1998). The surrounding starburst can thereby obscure the active
nucleus.

\begin{figure}
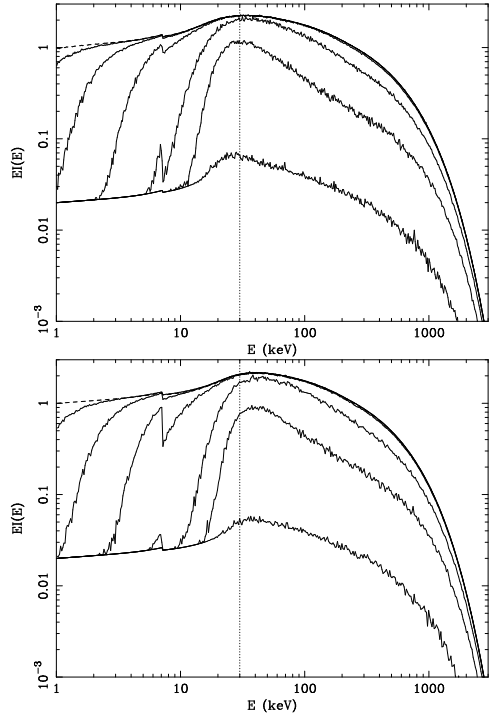

\centerline{\psfig{figure=mcspectraFe1.ps,width=0.5\textwidth,angle=270}}
\centerline{\psfig{figure=mcspectraFe5.ps,width=0.5\textwidth,angle=270}}
\caption{Monte-Carlo simulations of of an accretion disc spectrum (a
power-law of unit energy index with cold reflection and an exponential
cutoff at 360~keV) propagated through a solar abundance spherical
cloud of column densities ranging from $10^{21.25}$ to
$10^{25.25}$~cm$^{-2}$ in steps of a factor 10. Note that the spectra
peak around 30~keV, indicated by the vertical line. The lower panel
shows the effect of increasing the iron abundance by 5, which causes
the peak to shift to 40--50~keV (from Wilman \& Fabian 1999).}
\end{figure}

The starburst should enhance the metallicity of the absorbing gas.
This makes a given mass of gas more efficient at absorbing X-rays and
indeed increases the effect of absorption before Compton
down-scattering comes into play. This is important in opening up the
parameter space for model-fitting of the XRB spectrum (Wilman \&
Fabian 1999).

Fuelling of the nucleus is an old problem (see e.g. Shlosman et al
1990). Although there may be lots of gas around the nucleus, angular
momentum may prevent it from rapidly accreting to the centre. In this
respect, a hot phase in the surrounding medium may be important, with
Bondi accretion from this phase being the dominant mechanism (see e.g.
Nulsen \& Fabian 1999). Angular momentum may be transported outward by
turbulence within such a hot phase, so allowing rapid accretion to
proceed.

\section{The mean luminosity of the distant AGN dominating the XRB
intensity}

Since the mass of the black hole in nearby galaxies appears to be
proportional to the spheroid mass, the mass function of black holes
must be similar in shape to the spheroid mass function. The mean black
hole mass is therefore that appropriate to an $L^*$ galaxy, or about
$3\times 10^8$~M$_{\odot}$. The Eddington limit of such a black hole
is about $3\times 10^{46}$~erg~s$^{-1}$ and its mass doubling
(Salpeter) time is about $3\times 10^7$~yr. If the typical mass black
hole has therefore grown from say a million solar mass one in $3\times
10^9$~yr (i.e. by $z\sim 2$), then we probably need $L>0.05
L_{Edd}\sim 10^{45}$erg~s$^{-1}$. This means that the typical growing
black hole was powerful and of quasar-like luminosity (indeed housing
a quasar at the centre).

Such an obscured powerful object would locally be classified as a
ULIRG (see Sanders \& Mirabel 1996), although the distant ones need
not be the same as the local ones, which are perhaps mainly fuelled by
mergers.

Of course it is possible that massive black holes grew inside galaxies
which themselves were merging back at $z\sim 2$. Nevertheless, unless
they were all assembled from smaller holes just before accretion
switched off, it is probable that they emit for a reasonable fraction
of the last doubling time as a single object.

\section{Obscuration in a growing, isothermal galaxy spheroid}

Consider an isothermal galaxy in which a significant fraction $f_c$ of
cooled gas remains as cold dusty clouds instead of rapidly forming
stars. At the centre a black hole grows by accretion from the
surrounding cold (and hot) gas. Assume that the nucleus also blows a
wind of velocity $v_w$ which has a power $L_w=\alpha L_{Edd}$.
Eventually the wind becomes powerful enough to blow away the
surrounding gas and so shut off the accretion and further growth to
the black hole and spheroid. The Magorrian et al (1998) black-hole --
spheroid mass relation can then be obtained (Silk \& Ress 1998; Fabian
1999; Blandford 1999).

The kinetic power of a wind at which it ejects cold gas of column
density $N_H$ from a spheroid is given by $$L_W\approx 2\pi G M_{sph}
m_p N_H v_w$$ or
$$L_w\approx f_c \sigma^4 v_w G^{-1},$$ where $\sigma$ is the velocity
dispersion within the spheroid.  (I have used a force argument here,
see Fabian 1999; Silk \& Rees 1998 use an energy argument to obtain a
limit of $\sigma^5/G$, which is a factor $\sigma/v_w$ smaller than the
above $L_w$.) Ejection occurs when $$M_{bh}\approx {{\sigma^4
\sigma_T}\over{4\pi G^2 m_p}}{v_w\over c}{f_c\over \alpha}.$$ Using
the Faber-Jackson relation for spheroids ($M_{sph}\propto \sigma^4$)
then yields, if ${v_w\over c}{f_c\over \alpha}\sim 1$
$$M_b\sim 0.005 M_{sph},$$ close to the Magorrian et al (1998) relation.

At that point the column density in to the accretion radius $N_H\sim
N_T=\sigma_T^{-1}$, so the growth is (just) Compton thick. The growth
of massive black holes is radiatively efficient, highly obscured and
gives rise to much of the XRB. It is also intimately linked with the
growth of galaxy spheroids, the main evolution of which is terminated
by a quasar wind. X-ray observations probe best the underlying
obscured nucleus at (rest frame) ebergies of about 30~keV. Indeed
X-rays are the best diagnostic of the black hole accretion history of
the Universe.

The optically bright quasar phase (from an outside observer's point of
view) follows over the next few million years as the accretion disc
around the black hole empties. The early phase as the wind clears the
gas away can be identified with BAL quasars. The central engine is
only revived after the quasar phase if a merger or other event brings
in sufficient low angular momentum gas to fuel it.

\begin{figure}
\centerline{\psfig{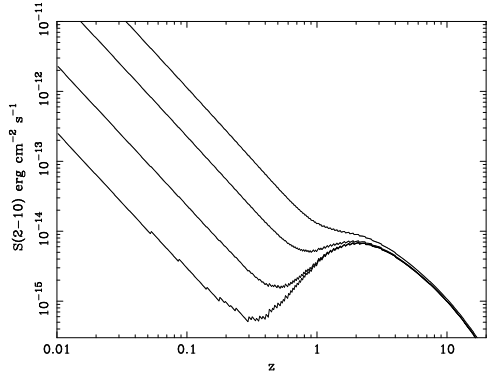}}
\caption{The observed 2--10~keV flux as a function of redshift from a
source of intrinsic (unabsorbed) 2--10~keV luminosity of
$10^{45}$~erg~s${-1}$ with a column density of $10^{24.5}$~cm$^{-2}$.
Scattering fractions (by thin ionized gas) of 5, 1, 0.1 and 0.01 per
cent are included (top to bottom). Note that the negatice K-correction
means that sources at $z\sim0.1, 0.8$ and 7 can have the same observed
2--10~keV flux. From Wilman \& Fabian (1999).}
\end{figure}

The prospects of testing the above scenario and absorption models of
the XRB are close at hand, with Chandra and XMM. They should detect
large numbers of faint, but powerful absorbed sources in the 3--10~keV
band, due to the negative K correction involved (see Fig. 3) and
identify them with luminous FIR/sub-mm--emitting young galaxy
spheroids.

\begin{acknowledgements}
I am grateful to Kazushi Iwasawa, Paul Nulsen and Richard Wilman for
continued collaboration and the organisers for an interesting
conference. The Royal Society is thanked for support.
\end{acknowledgements}

\end{document}